\documentclass{article}
\usepackage{spconf,amsmath,graphicx}
\usepackage{amsfonts}
\newcommand{\bftab}{\fontseries{b}\selectfont}
\usepackage{booktabs}
\usepackage{multirow}
\usepackage{siunitx,etoolbox}
\usepackage[keeplastbox]{flushend}
\usepackage[utf8]{inputenc}
\usepackage{pgfplots}
\DeclareUnicodeCharacter{2212}{−}
\usepgfplotslibrary{groupplots,dateplot}
\usetikzlibrary{patterns,shapes.arrows}
\pgfplotsset{compat=newest}
\usepackage[all=normal,floats=tight,charwidths=tight]{savetrees}
\usepackage{color,soul}
\usepackage{hyperref}
\usepackage{cite}

\title{Variational autoencoder for speech enhancement with a noise-aware encoder}
\name{Huajian Fang$^{1,2}$, Guillaume Carbajal$^1$,  Stefan Wermter$^{2}$, Timo Gerkmann$^1$}
\address{
  $^1$Signal Processing (SP), Universität Hamburg, Germany\\
  $^2$Knowledge Technology (WTM), Universität Hamburg, Germany\\
  \{fang, carbajal, wermter, gerkmann\}@informatik.uni-hamburg.de}

\begin{document}
\ninept

\maketitle
\begin{abstract}
Recently, a generative variational autoencoder (VAE) has been proposed for speech enhancement to model speech statistics. However, this approach only uses clean speech in the training phase, making the estimation particularly sensitive to noise presence, especially in low signal-to-noise ratios (SNRs). To increase the robustness of the VAE, we propose to include noise information in the training phase by using a \emph{noise-aware encoder} trained on noisy-clean speech pairs. We evaluate our approach on real recordings of different noisy environments and acoustic conditions using two different noise datasets. We show that our proposed noise-aware VAE outperforms the standard VAE in terms of overall distortion without increasing the number of model parameters. At the same time, we demonstrate that our model is capable of generalizing to unseen noise conditions better than a supervised feedforward deep neural network (DNN). Furthermore, we demonstrate the robustness of the model performance to a reduction of the noisy-clean speech training data size.
\end{abstract}

\section{Introduction}
\label{sec:intro}
Speech enhancement refers to the problem of extracting a target speech signal from a noisy mixture in order to enhance the quality and intelligibility of the speech. This task is of particular interest for applications like speech recognition and hearing aids. Single-channel speech enhancement is a challenging task, especially at low signal-to-noise ratios (SNRs).

Speech enhancement typically requires the statistical estimation of the noise and speech power spectral densities (PSDs) \cite{gerkmann2011unbiased, noisereducitonsurvey}. Non-negative matrix factorization (NMF) is a popular choice for PSD estimation \cite{lee2001algorithms,fevotte2009nonnegative,mohammadiha2011new,sawada2013multichannel}. However, underlying linearity assumptions limit the performance when modeling complex high-dimensional data. In contrast, speech enhancement based on non-linear deep neural networks (DNNs) has shown better modeling capacity. Common approaches focus on inferring a time-frequency mask in a supervised manner \cite{wang2018supervised}. However, to generalize to unseen noise conditions, DNNs require a large number of pairs of noisy and clean speech in various acoustic conditions \cite{rehr2019analysis}.

Recently, there has been an increasing interest in generative models, such as generative adversarial networks (GANs) \cite{goodfellow2014generative} and variational autoencoders (VAEs) \cite{kinmavae, rezende2014stochastic}. The generative VAE is a probabilistic model widely used for learning latent representations of a probabilistic distribution. The VAE features a similar architecture as a classical autoencoder with an encoder and a decoder, but its latent space differs by being regularized to follow a standard Gaussian distribution. Moreover, the VAE has been extended to deep conditional generative models for effectively performing probabilistic inference \cite{kingma2014semi, sohn2015learning}. VAEs have been applied to speech enhancement in both single-channel and multi-channel scenarios \cite{bandovae, simonvae, multichannelvae}. They have been used to model the speech statistics by training on clean speech spectra only. However, because no noise information is involved in its training phase, the encoder of the standard VAE is sensitive to noise. In low SNRs, this noise-sensitivity results in the erroneous estimation of latent variables and thus in inappropriately generated speech coefficients and a reduced performance.

In this work, inspired by conditional VAEs and its application to image segmentation \cite{kingma2014semi, sohn2015learning, kohl2018probabilistic}, to increase noise robustness, we propose to replace the encoder of the VAE by a \emph{noise-aware encoder}. To learn this encoder, the VAE is first trained on clean speech spectra only, and then, given noisy speech, the proposed noise-aware encoder is trained in a supervised fashion to make its latent space as close as possible to that of the first speech-only trained encoder. For our analyses we rely on the VAE-NMF speech enhancement framework \cite{bandovae, simonvae}, which  uses NMF to model the noise PSD.  We show that the proposed encoder is more robust to noise presence and improves speech estimation without increasing the number of model parameters. The method also shows robustness to unseen noise conditions by evaluating on real recordings from different noise datasets. Finally, we illustrate that already a small amount of noisy-clean speech data can lead to improvements in overall distortion.

In section~\ref{problemformulation}, we introduce problem settings and notations, as well as the framework of the VAE-based speech model and the noise model developed on the NMF. In section~\ref{sec:proposedVAE}, we introduce details about the proposed noise-aware VAE. After showing the experiment settings in section~\ref{experimentalsettings}, we present experimental evaluation results and conclusions in section~\ref{sec:results} and section~\ref{sec:conclusion}.

\section{Problem formulation}
\label{problemformulation}

\subsection{Mixture model}
\label{mixturemodel}
In our work, we employ an additive signal model, where a
noisy mixture is seen as a
superposition of clean speech 
and additive noise. 
In the short-time Fourier transform (STFT) domain, it shows as
\begin{equation}
x_{ft} = s_{ft} + n_{ft},
\label{eqn:timemodel}
\end{equation}
where $x_{ft}$, $s_{ft}$, and $n_{ft}$ represent each time-frequency coefficient in spectra of noisy mixture $X \in \mathbb{C}^{F \times T} $, speech $S \in \mathbb{C}^{F \times T}$, and noise $N \in \mathbb{C}^{F \times T} $ respectively. $F$ denotes the number of frequency bins, $T$ represents the number of time frames, which are indexed by $f$ and $t$, respectively. The speech and noise spectra are assumed to be mutually independent complex Gaussian distributions with zero-mean, i.e., $s_{ft} \sim \mathcal{N}_\mathbb{C}(0,\,\sigma^{2}_{s,ft}),
n_{ft} \sim \mathcal{N}_\mathbb{C}(0,\,\sigma^{2}_{n,ft})$
where $\sigma^{2}_{s,ft}$,  $\sigma^{2}_{n,ft}$ represent the variances of speech and noise. The PSD of signals is characterized by the parameter variance under the local stationary assumption \cite{localgaussian}. 

Furthermore, to provide an increased robustness to the loudness of the audio utterances, a time-dependent and frequency-independent gain $g_t$ is introduced  \cite{simonvae}. Eventually, this modifies the additive mixture model in~\eqref{eqn:timemodel} to
\begin{equation}
x_{ft} = \sqrt{g_t}s_{ft} + n_{ft}.
\label{eqn:gaintimemodel}
\end{equation}
Given the observed noisy mixture which follows a complex Gaussian distribution as $x_{ft} \sim \mathcal{N}_\mathbb{C}(0,\,g_t\sigma^{2}_{s,ft}+\sigma^{2}_{n,ft})$,
the desired speech can be extracted by separately modeling the speech and noise variances. 

\label{enhancementmodel}
\subsection{Speech model}
\label{sec:speechmodel}
For the VAE-based speech model, a frame-wise $D$-dimensional latent variable $z_t \in \mathbb{R}^{D} $ is defined, and an $F$-dimensional speech frame $s_{t}$ is assumed to be sampled from the conditional likelihood distribution $p_\theta(s_t|z_t)$. This is achieved by the decoder of VAE, also called the generative model. The variable $\theta$ here indicates the parameters of the decoder network. $\hat{\sigma}^{2}_{s}: \mathbb{R}^{D} \to \mathbb{R}_+^F$ denotes the nonlinear function from the latent space to the reconstructed signal given by the generative model of the VAE.

The VAE provides a principled method to jointly learn latent variables and the inference model \cite{kinmavae}. Following a Bayesian framework, this requires to approximate the intractable true posterior distribution $p(z_t|s_t)$. In the VAE, the encoder, also called the inference model, is used to approximate the true posterior, denoted as $q_\phi(z_t|s_t)$. The variable $\phi$ here indicates the parameters of the encoder network. $\hat{\mu}_d:  \mathbb{R}^{F}_+ \to \mathbb{R}^D $, $\hat{\sigma}_d^2: \mathbb{R}^{F}_+ \to \mathbb{R}^D_+$ indicate the nonlinear mapping of the neural network given by the inference model of the VAE.  Under stochastic gradient descent, the generative model's parameters $\theta$ and the inference model's parameters $\phi$ are jointly optimized by maximizing variational lower bound, given by
\begin{equation}
\label{eqn:elbo}
\begin{split}
\log p(S) \geq & - \sum_t \mathbb{KL}[q_\phi(z_t|s_t)||p(z_t))] \\ 
& + \sum_t\mathbb{E}_{q_\phi(z_t|s_t)}[\log p_\theta(s_t|z_t)]. 
\end{split}
\end{equation}
The quantity $p(z_t)$ represents the prior distribution of the $D$-dimen\-sional variable $z_t$, and $\mathbb{KL}$ indicates Kullback-Leibler divergence. The prior of the latent variables is defined as a zero-mean isotropic multivariate Gaussian $z_{t} \sim \mathcal{N}(\mathbf{0},\, \mathbf{I})$ as in \cite{kinmavae}. The first term in the objective function~\eqref{eqn:elbo} refers to the regularization error in the latent space to ensure meaningful latent variables, and the second term is the reconstruction error.
 
\begin{figure}[tb!]
\begin{minipage}[b]{1.0\linewidth}
  \centering
  \centerline{\includegraphics[width=7.6cm]{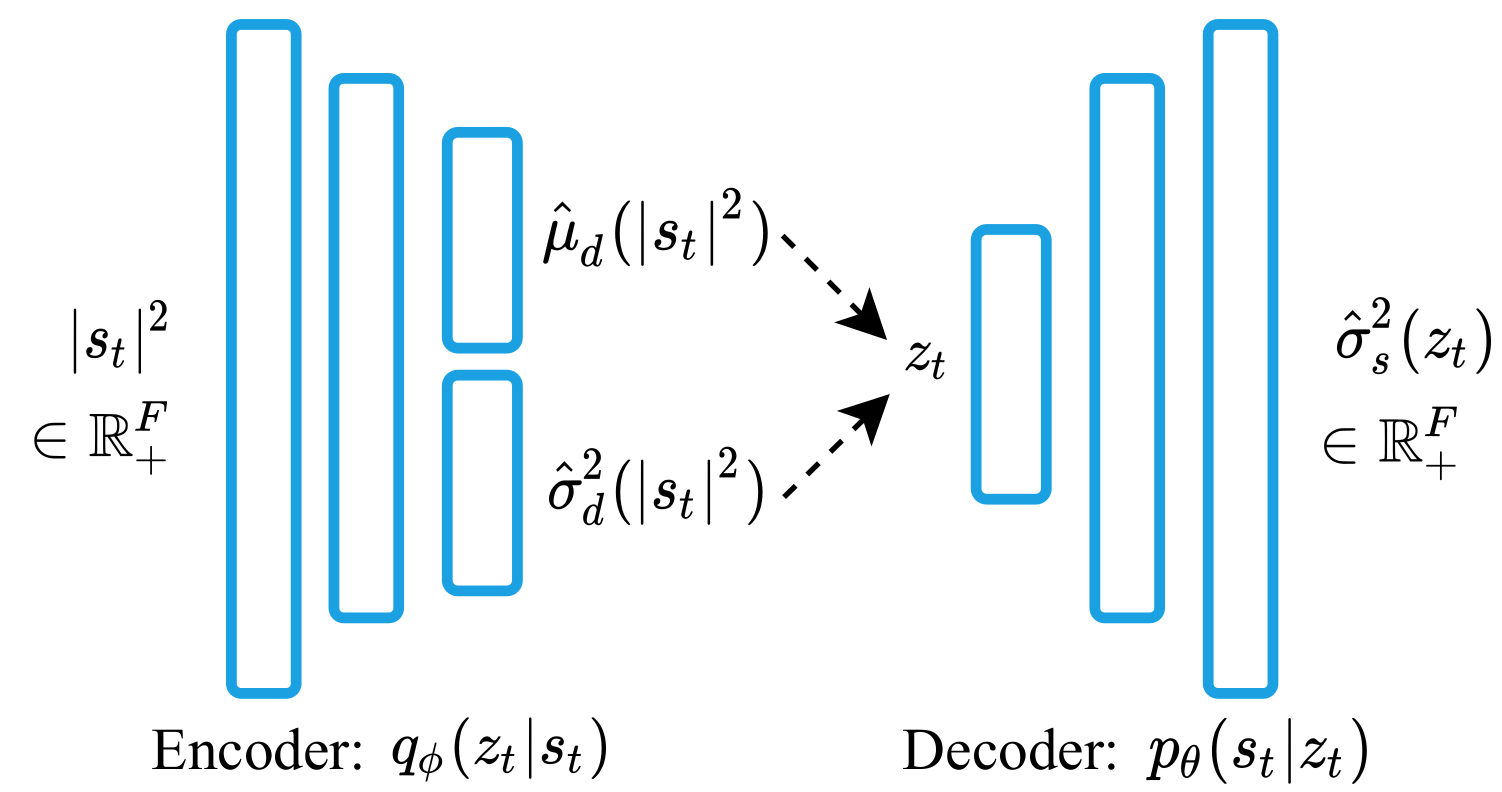}}
\end{minipage}
\caption{The generative model and inference model of the adopted VAE. The dashed line here indicates the sampling process.}
\label{fig:vae}
\end{figure}

As shown in Fig.~\ref{fig:vae}, the VAE is trained on the periodograms  of clean speech $\lvert s_t \rvert^2$ \cite{bandovae,simonvae}. During testing, the estimates of the clean speech power spectra $\hat{\sigma}^2_{s}(z_t)$ are expected to be generated from latent variables learnt from the noisy periodograms $\lvert x_t \rvert^2 \in \mathbb{R}^{F}_+$. Note that a robust estimation of latent variables that represents the clean speech statistics plays a crucial role in the generative process.  

\subsection{Noise model}
\label{sec:noisemodel}
NMF tries to find an optimal approximation to an input matrix by a dictionary matrix containing basis functions weighted by a coefficients matrix \cite{lee2001algorithms}. Here NMF is used to model the noise variance \cite{bandovae,simonvae}. The variance of noise $\sigma^{2}_{n}$ is approximated by a multiplication of the dictionary matrix $W \in \mathbb{R}_+^{F \times K}$ and the coefficients matrix $H \in \mathbb{R}_+^{K \times T}$, computed as
\begin{equation}
\label{eqn:nmf}
\sigma^{2}_{n}  = WH = \sum_{ft} \sum_k w_{fk}h_{kt},
\end{equation}
where $K$ indicates the rank of the noise model indexed by $k$. $w_{fk}$ and $h_{kt}$ are elements from $W$ and $H$ respectively at the corresponding row and column indexed by $f$, $k$, and $t$.

\subsection{Clean speech inference}
\label{speechinference}
By modeling speech and noise with VAE and NMF respectively, the distribution of the noisy mixture can be represented as
\begin{equation}
x_{ft} \sim \mathcal{N}_\mathbb{C}(0, g_t\hat{\sigma}^{2}_{s,f}(z_t) + \sum_k w_{fk}h_{kt}),  
\end{equation}
where $\hat{\sigma}^{2}_{s,f}: \mathbb{R}^{D} \to \mathbb{R}_+$ denotes the nonlinear function $\hat{\sigma}^2_s$ for $f$-th frequency bin. Given the noisy mixture as an observation, the Monte Carlo expectation-maximization (MCEM) algorithm is utilized to estimate the NMF parameters and the gain factor \cite{wei1990monte, simonvae}. The sampling strategy is based on the Metropolis-Hastings algorithm \cite{robert2013monte}. 
The clean speech can be extracted from a noisy mixture in the time-frequency domain by constructing a Wiener filter denoted by $\hat{m}_{ft}$, given as
\begin{equation}
    \hat{m}_{ft} = \frac{\hat{\sigma}^{2}_{s,f}(z_t) }{g_t\hat{\sigma}^{2}_{s,f}(z_t)  + \sum_k w_{fk}h_{kt}}.
\end{equation}

Although modeling speech with a VAE can be achieved by training solely on clean speech data, using it for speech enhancement is another matter since gaining robustness to noise is difficult without including noise samples in the training data and the model. However, the standard VAE does not allow for including noise at the training phase.

\section{Noise-aware VAE}
\label{sec:proposedVAE}

Instead of using the encoder trained on the clean speech signals, we propose a noise-aware VAE that can improve the robustness of the encoder against noise presence. For a generative process, it is difficult or even impossible to derive the optimal mapping between latent variables and targets. However, we argue that it might be relevant to make latent variables estimated from noisy mixtures as close as possible to the ones inferred from the corresponding clean speech. 

\begin{figure}[t]

\begin{minipage}[b]{.48\linewidth}
  \centering
  \centerline{\includegraphics[width=4.8cm]{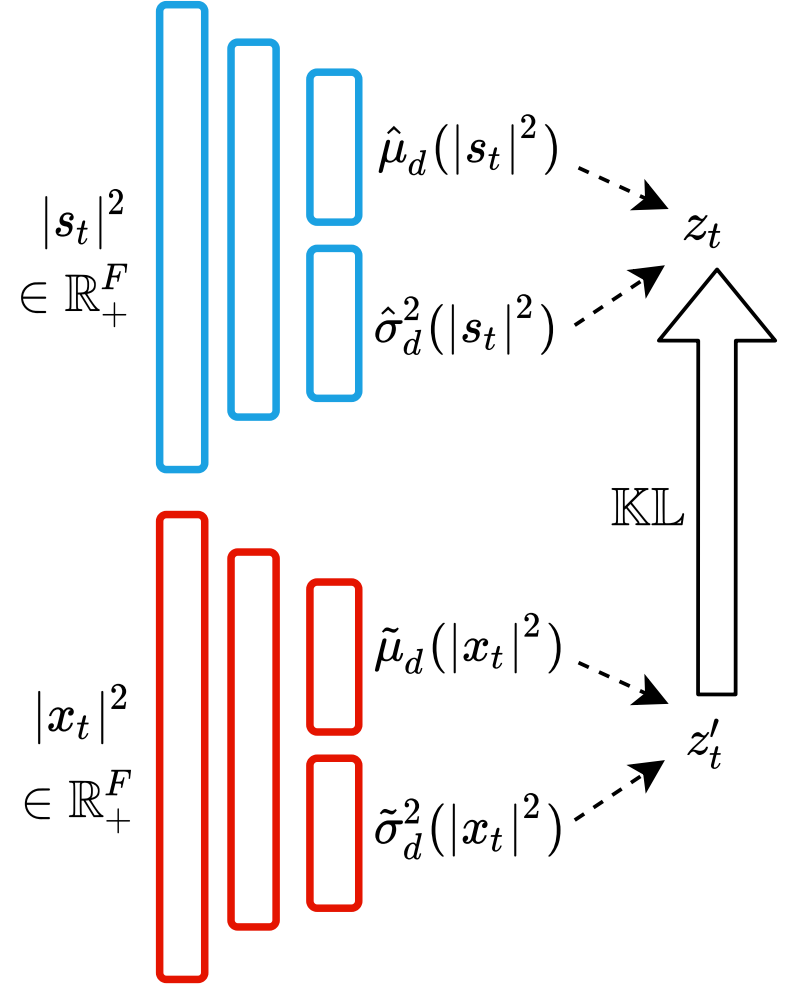}}
  \centerline{(a)}\medskip
\end{minipage}
\hfill
\begin{minipage}[b]{0.48\linewidth}
  \centering
  \centerline{\includegraphics[width=3.6cm]{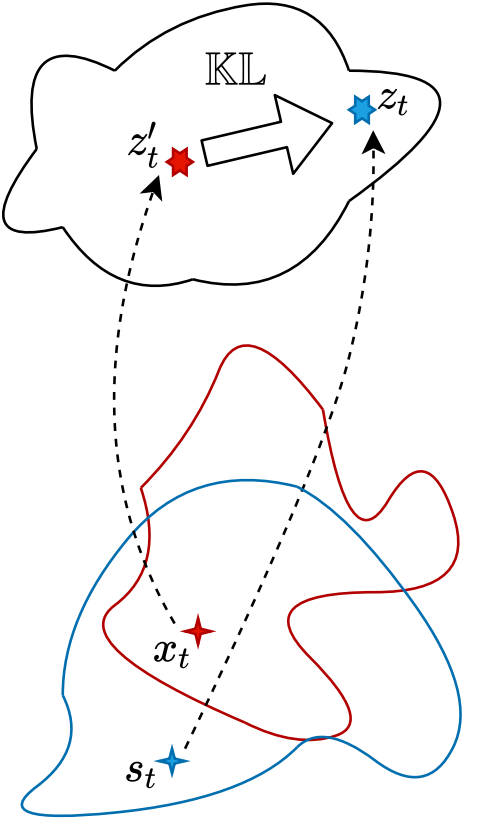}}
  \centerline{(b)}\medskip
\end{minipage}
\caption{The proposed architecture for minimizing divergence between latent variables. The constraint in the latent space is shown in (a), and its graphic explanation given in (b).}
\label{fig:pmodel}
\end{figure}

To obtain the noise-aware VAE based on this assumption, we propose a two-step learning algorithm, which learns a non-linear mapping from the noisy signals to latent variables that represent the clean speech statistics. We first train a VAE using Equation~\eqref{eqn:elbo} to learn a regularized latent space over the clean speech signals. The noise-aware encoder is then proposed to approximate the probability $q_\gamma(z^\prime_t|x_t)$ to output D-dimensional latent variables $z^\prime_t \in \mathbb{R}^{D}$ conditioned on the noisy mixture $x_t$. It is also assumed that the conditional probability $q_\gamma(z^\prime_t|x_t)$ follows a standard Gaussian distribution. The variable $\gamma$ indicates the parameters of the new encoder. 
Finally, the distance of $z^\prime_t$ obtained from noisy speech to the latent variables $z_t$ inferred form the corresponding clean speech is minimized based on the Kullback–Leibler divergence as shown in Fig.~\ref{fig:pmodel} (a), given by
\begin{align}
\label{kl}
\mathcal{L}(\gamma) &= \sum_{t} \mathbb{KL}(q_\phi(z_t|s_t)||q_\gamma ^\prime(z_t^\prime|x_t)) \\
\begin{split}
&= \sum_{t,d} 
 \Big\{\, \frac{1}{2}
\log \frac{{\widetilde{\sigma}_d^{2}(|x_t|^2)}}{{\hat{\sigma}^2_{d}(|s_t|^2)}} -\frac{1}{2} \\
&\qquad + \frac{\hat{\sigma}^2_{d}(|s_t|^2) + (\hat{\mu}_{d}(|s_t|^2) -\widetilde{\mu}_{d}(|x_t|^2))^2}{2\widetilde{\sigma}_d^{2}(|x_t|^2)}\, \Big\}
\end{split}
\end{align}
where $\widetilde{\mu}_{d}:\mathbb{R}^F_{+} \to \mathbb{R}^D$ and $\widetilde{\sigma}_d^2:\mathbb{R}^F_{+} \to \mathbb{R}_+^D$ represents the nonlinear mapping of the neural networks for the mean and variance of the posterior Gaussian distribution for the variable $z^\prime_t$. The parameters of the new inference model $\gamma$ are optimized by minimizing the cost function using stochastic gradient descent algorithms.
In this way, we combine unsupervised learning of the speech characteristics by the VAE and supervised learning using the pairs of noisy-clean speech signals.

Eventually, as graphically shown in Fig.~\ref{fig:pmodel} (b), by introducing this cost function in the latent space, the latent variables $z^\prime_t$ estimated from the noisy mixture $x_t$ is pulled towards $z_t$ estimated from the corresponding clean speech $s_t$. The dashed lines here indicate the nonlinear mapping from the signal space to the latent space, and different colors indicate two mapping pairs. At the inference stage, the noise-aware inference model is used to replace the standard speech-based encoder. The decoder of the VAE remains unchanged. 

\begin{table*}[tb]
\centering
\begin{tabular}{|c||c||c|c|c|c|c|}
 \hline
 SNR  & Average & -10 dB & -5 dB & 0 dB & 5 dB & 10 dB\\
 \hline
 Unprocessed & -0.04 $\pm$ 0.44 & -10.02 $\pm$ 0.03 & -5.03 $\pm$ 0.01 & -0.03 $\pm$ 0.01  & 4.95 $\pm$ 0.01 & 9.90 $\pm$ 0.02 \\
 \hline
 DNN-WF
 & 6.92 $\pm$ 0.42 & -1.96 $\pm$ 0.66 & 3.43 $\pm$ 0.53 & 7.25 $\pm$ 0.42 & 11.58 $\pm$ 0.38 & 14.25 $\pm$ 0.34  \\
 
 VAE 
 &  6.72 $\pm$ 0.43  & -1.92 $\pm$ 0.75 & 2.99 $\pm$ 0.59 & 6.89 $\pm$ 0.49 & 11.43 $\pm$ 0.42 & 14.14 $\pm$ 0.37 \\
 
 proposed NA-VAE
 & \bftab 7.29 $\pm$ 0.43  &  \bftab -1.00 $\pm$ 0.78  & \bftab 3.64 $\pm$ 0.59 & \bftab 7.30 $\pm$ 0.50 &\bftab 11.85 $\pm$ 0.42 & \bftab 14.57 $\pm$ 0.39  \\
 \hline
\end{tabular}
  \caption{Performance comparison in SI-SDR on 5 different SNR conditions trained and evaluated on different subsets of the QUT-NOISE dataset (4 noise types). Values of SI-SDR are given in mean $\pm$ confidence interval (95\% confidence) over all utterances of the evaluation dataset with unit dB. NA-VAE refers to the proposed noise-aware VAE.}
  \label{quteval}
\end{table*}

\begin{table*}[tb]
\centering
\begin{tabular}{|c||c||c|c|c|c|c|}
 \hline
 SNR  &  Average & -10 dB & -5 dB & 0 dB & 5 dB & 10 dB\\
 \hline
 Unprocessed & -0.04 $\pm$ 0.44 & -10.01 $\pm$ 0.01 & -5.02 $\pm$ 0.01 & -0.03 $\pm$ 0.01 & 4.95 $\pm$ 0.01 & 9.90 $\pm$ 0.02 \\
 \hline
 DNN-WF
 & 2.93 $\pm$ 0.45 & -7.38 $\pm$ 0.38 & -1.65 $\pm$ 0.26 & 3.25 $\pm$ 0.24 & 8.07 $\pm$ 0.22 & 12.34 $\pm$ 0.21 \\
 
 VAE 
 &  11.44 $\pm$ 0.54  & 2.74 $\pm$ 1.20  & 7.90 $\pm$ 1.07  & 12.27 $\pm$ 0.90 & 15.27 $\pm$ 0.72 & 19.02 $\pm$ 0.68 \\
 
 proposed NA-VAE
 & \bftab 11.88 $\pm$ 0.52  &  \bftab 3.45 $\pm$ 1.10 & \bftab 8.60 $\pm$ 1.03 & \bftab 12.70 $\pm$ 0.89 &\bftab 15.63 $\pm$ 0.71 &\bftab 19.06 $\pm$ 0.67  \\
 \hline
\end{tabular}
  \caption{Performance comparison in SI-SDR on 5 different SNR conditions trained on the QUT-NOISE dataset and evaluated on the DEMAND dataset (12 noise types, completely unseen noise conditions). Values of SI-SDR are given in mean $\pm$ confidence interval (95\% confidence) over all utterances of the evaluation dataset with unit dB.}
  \label{demandeval}
\end{table*}

\section{Experimental settings}
\label{experimentalsettings}
\subsection{Datasets}
We evaluate the performance of the proposed model by using signals from the speech dataset Wall Street Journal (WSJ0)\cite{garofolo1993csr}, and the noise databases QUT-NOISE \cite{dean2010qut} and DEMAND \cite{demandvincent}. QUT-NOISE is used in constructing datasets of both training and evaluation using 4 noise types "cafe", "car", "home", and "street" recorded in unique locations. DEMAND is introduced as another evaluation dataset corresponding to completely unseen noise conditions in the training set, and the noise signals are randomly sampled from recordings of 12 noise types in the categories "domestic", "public", "street", and "transportation".

To train the noise-aware encoder, around 25 hours of speech samples are chosen from WSJ0 and mixed with the sampled noise signals at a SNR randomly chosen from the range of -5 dB to 5 dB with a gap of 1 dB. Two speaker-independent evaluation datasets each containing around 2.3 hours of 1000 noisy samples are created by mixing the speech and noise signals at SNRs of -10 dB, -5 dB, 0 dB, 5 dB, and 10 dB.

\subsection{Baselines}
We show evaluation results by comparing the proposed noise-aware  VAE  to the standard  VAE,  and a  fully-connected  DNN model. The DNN model outputs a Wiener filter based on a  mean square error cost function \cite{maskapproximationmse}, referred to as DNN-WF. The standard VAE is trained on the same amount of the clean speech signals that are not mixed with the noise signals, while the supervised DNN-WF is trained on the same dataset as the noise-aware encoder.

\subsection{Hyperparameters}
All signals are sampled at 16 kHz. The signal is transformed into the STFT domain with a sine window of length 1024 ($F=513$) and a 25\% hop size. Global normalization to zero mean and unit standard deviation is employed for training the noise-aware encoder, since Kullback–Leibler divergence is scale-dependent.
The rank of NMF is chosen to be $K=8$ when modeling noise, and its composing matrices $W$ and $H$ are randomly initialized. The parameters of MCEM algorithm follow the setting in \cite{simonvae}.

The VAE is comprised of an encoder and a decoder both with two feedforward hidden layers of 128 units. The hyperbolic tangent activation function is applied to all hidden layers, except the output layer. The dimension of the latent space $L$ is fixed at 16. The noise-aware encoder has the same structure as the speech-based encoder of the standard VAE. The fully supervised DNN-WF contains 5 hidden layers, each with 128 units, and its architecture is built to contain a similar number of parameters as our VAE model. No temporal information is considered in DNN-WF, which is consistent with the non-sequential characteristic of the VAE. We apply the ReLU activation function to all hidden layers, and the sigmoid function is put on the output layer to ensure the estimate of the Wiener filter mask lies in the range $[0, 1]$. The parameters $\theta$ and $\phi $ of the VAE are optimized by Adam \cite{adam} with a learning rate of 1e-3, and the parameters $\gamma$ of the noise-aware encoder  with a learning rate of 1e-4.

\subsection{Evaluation metrics}
To show the enhancement performance, we employ scale-invariant signal-to-distortion ratio (SI-SDR) in decibel (dB) \cite{sisdr} to measure the overall distortion, which takes both noise reduction and artifacts into account.

\section{Results and discussions}
\label{sec:results}

\begin{figure}[tb!]
\begin{minipage}[b]{1.0\linewidth}
  \centering
  \centerline{\includegraphics[width=7.0cm]{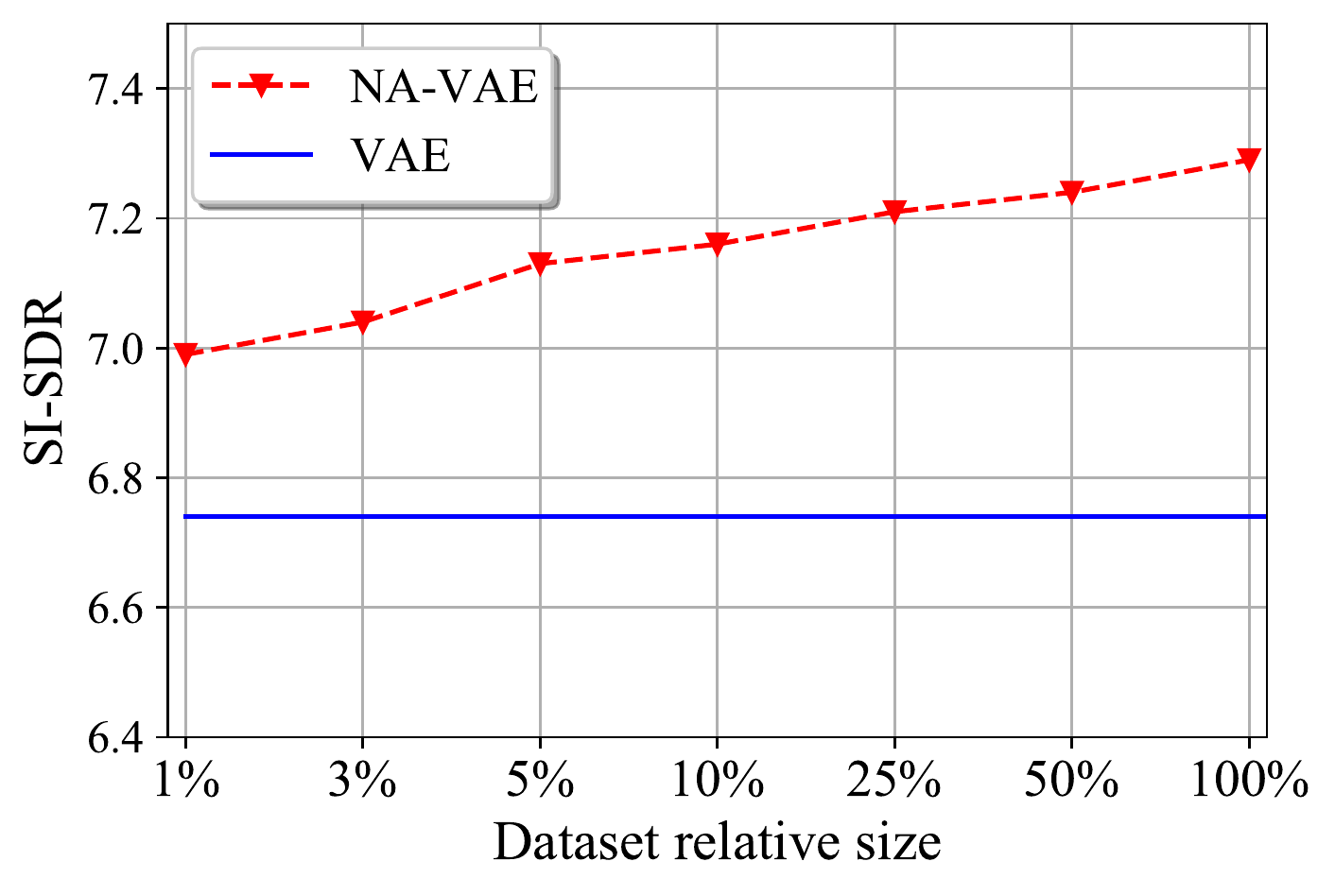}}
\end{minipage}
\caption{Influence of the amount of noisy-clean speech training data on SI-SDR improvements for both VAE models, averaged over all noise conditions.}
\label{estoi}
\label{fig:amountdata}
\end{figure}

\subsection{Performance evaluation}
As can be seen from the results in Table~\ref{quteval} which presents results trained and evaluated on different subsets of QUT-NOISE, the proposed noise-aware VAE outperforms the standard VAE in terms of overall distortion in all SNR scenarios, and the SI-SDR improvements are more evident at low SNR conditions. For example, the noise-aware VAE outperforms the baseline VAE by nearly 1 dB at an input SNR of -10 dB. Table~\ref{quteval} also shows that the DNN-WF performs better than the plain VAE, which implies that appropriate prior noise information is beneficial. In Table~\ref{demandeval}, which shows the evaluation performed on the DEMAND database while training is still conducted on QUT-NOISE, we see that the fully connected DNN-WF performs significantly worse than the other models. This was expected as we now test on a different more diverse dataset with 12 noise types instead of only 4. The supervised DNN-WF can not transfer the denoising capability to unseen noise types implying that inappropriate prior noise information may even deteriorate performance \cite{bandovae, rehr2019analysis}. However, the proposed noise-aware VAE can still outperform VAE in all SNR conditions, which suggests that the proposed method of improving latent variables in the latent space under this configuration is more capable of generalizing to unseen noise scenarios. Informal listening confirms the SI-SDR results especially for Table~\ref{quteval}, while the improvements reported in Table~\ref{demandeval} are relatively subtle. Audio examples are available online \footnote{\url{https://uhh.de/inf-sp-navae2021}}.

\subsection{Analysis of the amount of training data}
We then look at the influence of the amount of noisy-clean speech training data for estimating the speech latent variable. To achieve this, we initialize the noise-aware encoder with the encoder parameters of the pre-trained standard VAE and then train the new encoder by randomly selecting 1\%, 3\%, 5\%, 10\%, 25\%, 50\% of the noisy-clean speech pairs constructed with the QUT-NOISE dataset. In Fig.~\ref{fig:amountdata}, it is shown that the performance can already be improved by using only a small percentage of the paired noisy-clean speech data. A value of more than 0.2 dB SI-SDR improvement can be observed with just 1\% of the total paired data. It can also be observed that increasing the number of data in the later stage leads to gradual improvements, which
may be due to the noise diversity already being largely represented in the small fraction of data used. The research can be extended by increasing the diversity of the noise types in the training phase. This ability of improving performance with only few labeled data shows potential in alleviating overfitting issues in supervised training strategies.

\section{CONCLUSION}
\label{sec:conclusion}
In this paper, we proposed a noise-aware encoding scheme to improve the robustness of the VAE encoder particularly in low SNRs. For this we incorporate noise information into the VAE encoder to enable a more accurate speech variance estimation based on improved latent variables. By constraining the latent space, the VAE with the proposed noise-aware encoder can learn a non-linear mapping from the noisy mixture to latent variables that represent the clean speech statistics. Our proposed VAE outperforms the standard VAE and a supervised DNN-based filter in SI-SDR. Experiments also showed the generalization ability to unseen noise scenarios by evaluating across different datasets. Moreover, we showed that we could improve the performance even with a small amount of noisy-clean speech data. For future work, our approach could also be integrated with deep generative models that combine temporal dependencies \cite{juliusstcn}.

\bibliographystyle{IEEEbib}
\bibliography{lib}

\end{document}